\documentclass[a4paper,fleqn,usenatbib]{mnras}

\usepackage{newtxtext,newtxmath}

\usepackage[T1]{fontenc}
\usepackage{ae,aecompl}

\usepackage{graphicx}	

\usepackage{amsmath}	
\usepackage{amssymb}	
\usepackage{epstopdf}	

\usepackage{soul}
\usepackage{pifont}

\usepackage{comment}

\usepackage{scalerel,tikz}
\usetikzlibrary{svg.path}
\definecolor{orcidlogocol}{HTML}{A6CE39}
\tikzset{orcidlogo/.pic={
 \fill[orcidlogocol] svg{M256,128c0,70.7-57.3,128-128,128C57.3,256,0,198.7,0,128C0,57.3,57.3,0,128,0C198.7,0,256,57.3,256,128z};
 \fill[white] svg{M86.3,186.2H70.9V79.1h15.4v48.4V186.2z}
 svg{M108.9,79.1h41.6c39.6,0,57,28.3,57,53.6c0,27.5-21.5,53.6-56.8,53.6h-41.8V79.1z M124.3,172.4h24.5c34.9,0,42.9-26.5,42.9-39.7c0-21.5-13.7-39.7-43.7-39.7h-23.7V172.4z}
 svg{M88.7,56.8c0,5.5-4.5,10.1-10.1,10.1c-5.6,0-10.1-4.6-10.1-10.1c0-5.6,4.5-10.1,10.1-10.1C84.2,46.7,88.7,51.3,88.7,56.8z};
}}
\newcommand\orcid[1]{\href{https://orcid.org/#1}{\mbox{\scalerel*{
\begin{tikzpicture}[yscale=-1,transform shape]
\pic{orcidlogo};
\end{tikzpicture}
}{|}}}}

\title[Emergence of SMBHs from Pop III.1 Seeds]{The Emergence and Ionizing Feedback of Pop III.1 Stars as Progenitors for Supermassive Black Holes}

\author[Sanati et al.]{Mahsa Sanati$^{1,2}$\thanks{E-mail: mahsa.sanati@physics.ox.ac.uk},
Jonathan C. Tan$^{2,3}$,
Julien Devriendt$^{1}$,
Adrianne Slyz$^{1}$,
Sergio Martin-Alvarez$^{4}$,
\newauthor 
Matteo la Torre $^{5}$,
Benjamin Keller $^{11}$,
Maya A. Petkova$^{2}$,
Pierluigi Monaco$^{5,8,9,10}$,
Vieri Cammelli$^{6}$,
\newauthor 
Jasbir Singh$^{7}$,
 Matthew Hayes $^{12}$
\\
$^{1}$Department of Physics, University of Oxford, Keble Road, Oxford OX1 3RH, UK\\
$^{2}$Department of Space, Earth \& Environment, Chalmers University of Technology, SE-412 96 Gothenburg, Sweden\\
$^{3}$Department of Astronomy, University of Virginia, Charlottesville, VA 22904-4235, USA\\
$^{4}$Kavli Institute for Particle Astrophysics \& Cosmology (KIPAC), Stanford University, Stanford, CA 94305, USA\\
$^5$ Dipartimento di Fisica, Sezione di Astronomia, Università degli Studi di Trieste, via G.B. Tiepolo 11, I-34131, Trieste, Italy\\
$^6$ Department of Physics, Informatics \& Mathematics, University of Modena \& Reggio Emilia, via G. Campi 213/A, 41125, Modena, Italy \\
$^7$ INAF - Osservatorio Astronomico di Brera, via Brera 20, I-20121 Milano, Italy\\
$^8$ INAF - Osservatorio Astronomico di Trieste, via G.B. Tiepolo 11, I-34131, Trieste, Italy\\
$^9$ INFN, Sezione di Trieste, Via Valerio 2, 34127 Trieste TS, Italy\\
$^{10}$ IFPU, Institute for Fundamental Physics of the Universe, via Beirut 2, 34151 Trieste, Italy\\
$^{11}$ Department of Physics and Materials Science, University of Memphis, Memphis, TN 38152, USA\\
$^{12}$ Stockholm University, Department of Astronomy and Oskar Klein Centre for Cosmoparticle Physics, AlbaNova University Centre, SE-10691 Stockholm, Sweden\\
}

\date{Accepted XXX. Received YYY; in original form ZZZ}

\pubyear{2025}

\begin{document}

\newcommand{\papertitle}[2]{\textbf{#1 \citep{#2}}.\newline}

\label{firstpage}
\pagerange{\pageref{firstpage}--\pageref{lastpage}}
\maketitle

\begin{abstract}
Recent observations by JWST reveal an unexpectedly abundant population of rapidly growing supermassive black holes (SMBHs) in the early Universe, underscoring the need for improved models for their origin and growth. 
Employing new full radiative transfer hydrodynamical simulations of galaxy formation, we investigate the local and intergalactic feedback of SMBH progenitors for the Population III.1 
scenario, i.e., efficient formation of supermassive stars from
pristine, undisturbed dark matter minihalos.
Our cosmological simulations capture 
the \textit{R-type} expansion phase of these Pop III.1 stars, with their H-ionizing photon luminosities of $\sim$$10^{53}\:{\rm s}^{-1}$ 
generating HII regions that extend deep into the intergalactic medium, reaching comoving radii of $r_\text{HII}\sim$$ 1\,\text{cMpc}$. We vary both the Pop III.1 ionization flux and cosmological formation environments, finding the former regulates their final $r_\text{HII}$, whereas the latter is more important in setting their formation redshift. 
We use the results from our radiation-hydrodynamics simulations to estimate the cosmic number density of SMBHs, $n_{\mathrm{SMBH}}$, expected from Pop III.1 progenitors. 
We find $n_{\mathrm{SMBH}}\sim$$10^{-1}\,\mathrm{cMpc}^{-3}$, consistent with the results inferred from recent observations of the local and high redshift universe.
Overall, this establishes Pop III.1 progenitors as viable candidates for the formation of the first SMBHs, and emphasises the importance of exploring heavy mass seed scenarios. 
\newline

\end{abstract}

\definecolor{brown}{rgb}{0.5, 0.3, 0.0}
\definecolor{orange}{rgb}{0.8, 0.5, 0.0}


\section{Introduction}
\label{sec:intro}

Observations of luminous quasars, powered by accretion onto supermassive black holes (SMBHs) with masses $\sim 10^9\:M_\odot$ at high redshifts ($z\gtrsim6$) \citep[e.g.,][]{Mortlock:2011, banados2018, Wang:2021} present a significant challenge to our understanding of their origin in the early universe.
%
Recent surveys with the James Webb Space Telescope (JWST) have been able to probe to fainter SMBHs with masses $\sim 10^6 - 10^8\:M_\odot$ and found a surprising abundance of such objects 
\citep[e.g.,][]{harikane_jwstnirspec_2023, Fan2023, maiolino_jades_2023, 2024ApJ...964...39G, 2024NatAs...8..126B, Kokorev:2024}.
%
Their number density is further constrained by a variability analysis of the Hubble Ultra Deep Field (HUDF), which provides an observational estimate of the co-moving number density of high-redshift ($6<z<9$) AGNs, finding $n_{\rm SMBH}\gtrsim 10^{-2}\,{\rm cMpc}^{-3}$ \citep{hayes_glimmers_2024,2025arXiv250117675C}.
The same analysis also yields a constraint on the local ($z<0.5$) number density of $n_{\rm SMBH}\gtrsim 3\times 10^{-2}\:{\rm cMpc}^{-3}$ \citep{2025arXiv250117675C}.

%
The existence of SMBHs at very high redshifts challenges models of ``light'' ($\sim100\:M_\odot$) seeding. In particular, there is not enough time to reach the observed masses if accretion is limited to the Eddington rate. 
In these models black hole seeds form from the expected remnants of standard Pop III stars \citep[e.g.,][]{McKee&Tan2008} or growth from the mergers of stars in dense star clusters via intermediate-mass black holes (IMBHs) \citep[e.g.,][]{2004ApJ...604..632G,2022MNRAS.512.6192S}. 
%
%
It has been suggested that super-Eddington accretion could allow growth from light seeds to the observed high-$z$ SMBH population \citep[e.g.,][]{2024A&A...686A.256L}, however, there is limited observational evidence for such accretion rates in quasar populations \citep[e.g.,][]{Trakhtenbrot2017,2022ApJS..263...42W,Fragione&Pacucci2023,2024ApJ...966...85Z}. On the theoretical side, continuous growth at super-Eddington (or even Eddington) rates is considered unlikely due to  feedback from Pop III star progenitors \citep[e.g.,][]{2005ApJ...628L...5O,johnson_aftermath_2007,2009ApJ...699L.125M} or star formation and AGN feedback during SMBH accretion. Furthermore, observations of IMBHs, i.e., in the mass range of $\sim 10^2 - 10^4\,M_\odot$, in the local universe remain rare \citep[e.g.,][]{2016PASA...33...54R,2020ARA&A..58..257G,2024arXiv241017087M}. 
Thus, while light seed scenarios remain attractive and could still explain a broader and less extreme population of black holes,
these considerations favor mechanisms that produce ``heavy'' seeds with initial black hole masses of $\sim 10^4$ to $10^5\,M_\odot$.

The ``Direct Collapse'' (DC) mechanism has been proposed to produce heavy-seed black holes via the accretion of atomically-cooling, metal-free gas onto a protostar located in the centre of a relatively massive $\sim 10^8\:M_\odot$ dark matter halo. However, to suppress molecular cooling and avoid fragmentation, such systems require a strong far-UV radiation field from nearby sources \citep[e.g.,][]{Bromm&Loeb2003, 2006MNRAS.370..289B}.  
One of the main challenges for the DC scenario, particularly if it is invoked to explain the entire population of SMBHs, is its apparent rarity when simulated in cosmological volumes. 
For example, \citet{2019Natur.566...85W} carried out radiation-hydrodynamical simulations and estimated the co-moving number density of direct collapse black holes (DCBHs) to be $\lesssim 10^{-6}\:{\rm Mpc}^{-3}$. This value is more than four orders of magnitude lower than the observed co-moving number densities of SMBHs at both low and high redshifts \citep[e.g.,][]{hayes_glimmers_2024,2025arXiv250117675C}. \citet{2025arXiv250200574O} explored variations of how far-UV feedback impacts DCBHs seeding 
and found 
co-moving number densities of $\sim 10^{-4}\:{\rm cMpc}^{-3}$, still more that two orders of magnitude below observational estimates. 
Further variants on the DC model that incorporate turbulent gas support have yielded similarly low DCBHs number densities, e.g., $\lesssim10^{-6}\:{\rm cMpc}^{-3}$ in the simulations of \citet{2022Natur.607...48L}.

%

To reach SMBH number densities closer to observational estimates, some models have been proposed in which the stringent conditions traditionally associated with heavy seed formation are relaxed \citep{2022MNRAS.511..616T,2023MNRAS.521.2845C,2025arXiv250200574O}. For instance, \citet{2025OJAp....8E..11M} demonstrate that a number density of $\sim 10^{-2}\,{\rm cMpc}^{-3}$ can be achieved when seeding all atomically-cooling halos (i.e., with virial temperature $T_{\rm vir}\gtrsim10^4\,{\rm K}$), provided they meet additional criteria such as sub-threshold metallicity ($<10^{-3}\,Z_\odot$), compactness parameter $\gamma>0.5$, and gas inflow rates exceeding $0.1\,M_\odot\,{\rm yr}^{-1}$ (evaluated at a radius of $20\,\rm{pc}$). However,  as discussed in their work, these conditions may not necessarily lead to monolithic DCBHs, but instead to fragmentation into dense star clusters.
As such, the efficiency of black hole seed formation in these environments, and the resulting seed mass spectrum, remain open questions.

An alternative model that produces heavy seeds is based on the formation of supermassive Population III.1 stars in locally isolated dark matter minihalos \citep[][see \citet{2024arXiv241201828T} for a review]{banik_formation_2019,singh_formation_2023,2025MNRAS.536..851C}. The process that allows protostars to grow to high masses and produce heavy seeds is injection of energy from dark matter annihilation \citep{spolyar_dark_2008,2008IAUS..255...24T,2009ApJ...692..574N,2009ApJ...693.1563F,2015ApJ...799..210R}. This energy injection allows the protostar to remain in a large, swollen state with a relatively cool photospheric temperature and thus low level of ionizing feedback. As a result, it may accrete a significant fraction of the baryonic content of the minihalo, i.e., up to $\sim 10^5\:M_\odot$. In the cosmological model for Pop III.1 seeding, only isolated, undisturbed minihalos undergo this evolution, since it requires slow baryonic contraction to adiabatically concentrate the dark matter density distribution around the forming star. On the other hand, minihalos that survive irradiation by ionizing photons have enhanced free electron abundances and thus increased amounts of $\rm H_2$ and HD molecular coolants, leading to fragmentation to several lower mass stars and thus limited adiabatic contraction of the dark matter density. These so-called Pop III.2 sources are estimated to have masses of only $\sim 10\,M_\odot$ \citep[e.g.,][]{2006MNRAS.366..247J} and thus would not be a significant seeding mechanism for SMBHs. 
%
As discussed by \citet{2024arXiv241201828T}, Pop III.1 stars may emit substantial fluxes of ionizing radiation during a phase of their stellar evolution prior to SMBH formation, thereby regulating the global abundance of SMBH seeds, $n_{\rm SMBH}$.
If the Pop III.1 star has a mass of $\sim 10^5\:M_\odot$ and exists in a configuration near the main sequence, then its rate of production of H-ionizing photons is expected to be $\sim 10^{53}\:{\rm s}^{-1}$. While the standard main sequence lifetime of such a star is relatively short, i.e., $\sim3\times 10^6\:$yr, this phase may be lengthened by WIMP annihilation heating. 

So far the Pop III.1 SMBH seeding scenario has been investigated only in idealized cosmological simulations based on the PINOCCHIO (PINpointing Orbit Crossing Collapsed HIerarchical Objects) code \citep{2002JMS....33....1M,2013MNRAS.433.2389M, 2017MNRAS.465.4658M}. 
This code uses a Lagrangian Perturbation Theory (LPT) method \citep{1991ApJ...382..377M, 1993MNRAS.264..375B, 1995MNRAS.276..115C} for generation of catalogs of dark matter halos. 
\citet{banik_formation_2019} and \citet{singh_formation_2023} presented a simulation of Pop III.1 seeding based on a constant value of isolation distance, $d_{\rm iso}$, defined as the distance to the nearest minihalo hosting a Pop III.1 star, with typical value of $d_{\rm iso}\sim 50$ to 100~kpc (proper distance). Values of $d_{\rm iso}$ at the smaller end of this range are needed to yield $n_{\rm SMBH}\simeq 10^{-1}\:{\rm Mpc}^{-3}$. \citet{2025MNRAS.536..851C} presented semi-analytic models of galaxy evolution and SMBH growth based on the PINOCCHIO halo merger trees, also finding a preference for $d_{\rm iso}\lesssim 75\:$kpc based on comparison with SMBH occupation fractions and mass functions. 
In this paper it was shown that matching some AGN and host galaxy properties at high redshifts is challenging. However these estimates are based on a set of parametric models of star formation and AGN accretion that have been tested from cosmic noon to the present age, but their validity is not guaranteed to hold at very high redshift. This limitation underscores the need for more realistic simulations, which motivates the work presented in our study. 

Separately from semi-analytical models, SMBH seeding has been explored in numerous
cosmological volume simulations that aim to reproduce statistical properties of galaxies. However, these typically lack sufficient resolution to capture small-scale processes and so SMBH seeding is usually implemented with simple threshold conditions, e.g., on dark matter halo mass \citep[e.g.,][]{vogelsberger2014,Schaye:2015} or gas properties \citep[e.g.,][]{2017MNRAS.470.1121T}. 
%
Zoom-in simulations \citep[e.g.,][]{costa_quenching_2018-1,2018MNRAS.473.4003B, 2019Natur.566...85W, 2022MNRAS.513.3768I, 2023MNRAS.520.5394W, 2025arXiv250408041F,2025MNRAS.537.2559H} have been employed as a complementary technique, offering higher resolution and improved physical fidelity within selected regions of interest.

Here, we present a high-resolution cosmological zoom-in simulation of Pop III.1 sources, incorporating physically motivated models. This allows us to quantify the isolation distance of Pop III.1 progenitor halos set by the ionizing feedback from the star, and to estimate the resulting number density of SMBHs. 
The paper is organized as follows. The numerical framework to generate and evolve our simulations is described in \S\ref{sec:method}. The results are presented in \S\ref{sec:res}.
Finally, a summary of our main conclusions and a discussion of the implications and limitations of this work are presented in \S\ref{sec:sum}. 

\section{Numerical methods and simulations}\label{sec:method}

We generate all the cosmological simulations studied in this work using the adaptive mesh refinement (AMR) code \textsc{Ramses} \citep{Teyssier2002}. Two types of simulations are run: dark matter only (DMO) and zoom-in radiative-magneto-hydrodynamical (RT-MHD) simulations. 
While the main goal of hydro simulations is to investigate the ionizing feedback of Pop III.1 stars and to estimate the isolation distance of seeding halos, the DMO simulations are intended to provide estimates for the number density of SMBHs.
The DMO runs are described in more details in \S\ref{sec:nSMB}, and the main features of the hydro simulations are outlined here.

We use the adaptive mesh refinement of \textsc{Ramses} to resolve dense and Jeans-unstable regions. 
When the total dark matter and baryonic mass within a grid cell exceeds $8$ times its initial value, or the grid cell size surpasses $4$ local Jeans lengths, the parent cell is split into $8$ equal child cells. The size of each cell $i$ on the grid is determined by the refinement level $l_i$ according to $\Delta x_{i} = 1/2^{l_i}\,L_{\mathrm{Box}}$. 
We set the maximum level of refinement to $21$, which corresponds to a minimum cell size of $3.6\,\mathrm{pc}$.
In this RT-MHD version, \textsc{Ramses} solves the evolution of gas on this octree grid, while simultaneously and self-consistently modeling radiative transfer \citep{rosdahl_ramses-rt_2013}, and constrained transport magneto-hydrodynamics \citep[][]{2006A&A...457..371F, 2006JCoPh.218...44T},  
in addition to treating baryonic physics, such as redshift-evolving and uniform UV heating, gas cooling, star formation, and stellar feedback. These features are described in more detail later in this section.

The initial conditions for all simulations are generated at redshift $z = 100$, using the \textsc{MUSIC} code \citep{2011MNRAS.415.2101H}, with cosmological parameters adopted from \citet{2020A&A...641A...6P}.
The hydro simulations are generated employing a cubic box of size $L_{\mathrm{Box}}=7.55\,\mathrm{cMpc}$ per side.
We identify the first collapsing dark matter minihalo, which later becomes the site of Pop III.1 star formation. 
At the time of Pop III.1 star formation, which occurs at $z=22.5$ in our fiducial model, this halo has a virial mass of $\simeq 4\times 10^6\,M_\odot$ and contains $300$ gravitationally bound dark matter particles. 
As we are primarily interested in accurately following the 
collapse and evolution of this halo and its proto-galaxy, we resolve this sub-volume with a zoom-in region. This allows for higher resolutions at a lower computational expense. 
To select this zoom region, we identify all particles that eventually reside within the target halo by redshift $z=0$, and refine all particles within a 3D ellipsoid around this halo to a dark matter mass resolution of $m_{\mathrm{DM}}\simeq10^{4}\,M_{\odot}$ in the initial conditions.
Outside the zoom region, the resolution is gradually degraded from level
$10$ to $9$ using the \textsc{MUSIC} code. Note, one resolution level $l$ corresponds to $N=(2^l)^3$ particles in the full cosmological box. The particle mass is thus decreased by a factor of eight between these two levels.

\textbf{Star formation:}
We adopt a magneto-thermo-turbulent star formation prescription to model the formation of stars. This approach has been previously introduced in the hydrodynamical framework by \citet{Kimm2017} and \citet{Trebitsch2017} and extended to include MHD by \citet{Martin-Alvarez2020}.
In this model, gas is eligible for conversion into stellar particles only in gas cells where gravitational collapse exceeds the combined support from magnetic fields, thermal pressure, and turbulence. Furthermore, we only allow star formation to take place in the highest level of refinement \citep{Rasera2006}. 

As indicated above, we designate the first collapsing dark matter minihalo as the site of formation for the Pop III.1 star. This star is formed once our magneto-thermo-turbulent star formation criterion is fulfilled inside this minihalo.
The delay between the collapse of the progenitor halo and the formation of the Pop III.1 star, which was not accounted for in the studies by \citet{banik_formation_2019}, \citet{singh_formation_2023}, and \citet{2025MNRAS.536..851C}, is important to consider when developing improved models of SMBH populations forming via the Pop III.1 scenario (Petkova et al., in prep.).

The Pop III.1 star is modeled as a single stellar particle with a mass of $10^5\,M_{\odot}$. This mass is extracted from the baryonic gas mass in this minihalo, and constitutes a large proportion of its total baryonic content. 
Any further stars forming in the simulations will be treated as standard stellar particles, with a minimum mass of $\sim2.4\times10^2\,M_\odot$ that represents individual stellar populations rather than single stars.
%
For these stars, the conversion of gas into stellar particles is assumed to occur at a rate 
following the general form of a Schmidt law \citep{Schmidt1959}:
\begin{equation} {\dot\rho_{\star}} = \epsilon_{\mathrm{ff}}\frac{\rho_{\mathrm{gas}}}{t_{\mathrm{ff}}}. \end{equation}
Here, gas in a cell with density $\rho_{\mathrm{gas}}$ undergoes star formation over the local free-fall timescale $t_{\mathrm{ff}}$.
In the \textsc{Ramses} implementation, a fraction of the gas mass in a given cell is converted into a stellar particle, with the efficiency $\epsilon_{\mathrm{ff}}$, which depends on the local properties of gas, following the multi-scale prescription of \citet{2011ApJ...730...40P} as implemented in \citep{Federrath2012}.


\textbf{Stellar radiation:}
We employ the \textsc{RamsesRT} implementation by \citet{rosdahl_ramses-rt_2013} and \citet{2015MNRAS.449.4380R} for simulating the injection, propagation, and interaction of radiation with the multi-phase gas. 
Due to our relatively high spatial resolution of $\Delta x \approx 3.6\,\mathrm{pc}$, we expect well-resolved escape of ionizing radiation both from the model galaxy and within its ISM \citep{2014ApJ...788..121K}.

In its radiation hydrodynamics implementation, \textsc{RamsesRT} employs a first-order Godunov method with the $\mathrm{M1}$ closure \citep{Levermore1984,Dubroca&Feugeas1999} for the Eddington tensor. 
By using an explicit solver for the radiative transport, the advection time-step $\Delta t$, and consequently the CPU time, scale inversely with the speed of light $c$ as $\Delta t<\Delta x/(3c)$. This constraint mandates a time step significantly shorter than the hydrodynamic time step, which is limited by the maximum velocity of the gas ($\sim1000\,\mathrm{km\:s^{-1}}$). 
To mitigate this constraint, we adopt the ``reduced speed of light'' approximation \citep{2001NewA....6..437G}. 
We set the reduced speed of light at $0.2\,c$, and allow the radiation solver to subcycle over the hydrodynamical time-stepping up to a maximum of $500$ steps. This adjustment proves adequate for modeling the propagation of ionization fronts through both the ISM of galaxies and the lower-density environment of the IGM \citep{rosdahl_ramses-rt_2013}.


%
The three radiation groups in this work are divided into spectral bins as 
\begin{equation}
    \mathrm{Photon\, group}= 
    \begin{cases}
    \mathrm{HI} & 13.6\,\mathrm{eV}<\epsilon_{\mathrm{photon}}<24.59\,\mathrm{eV}\\
    \mathrm{HeI} & 24.59.6\,\mathrm{eV}<\epsilon_{\mathrm{photon}}<54.42\,\mathrm{eV}\\
    \mathrm{HeII} & \epsilon_{\mathrm{photon}}>54.42\,\mathrm{eV}\\
    \end{cases}
\end{equation}
%
%
In the simulations of this paper, only the stellar particles, including the Pop III.1 star, are sources of ionizing radiation, i.e., AGN feedback is not included. A detailed exploration of different methods of treating AGN feedback is deferred to a future paper in this series.
For the luminosity of normal stellar particles we use the Binary Population and Spectral Synthesis (\textsc{bpassv2.0}) model \citep{2008MNRAS.384.1109E, 2016MNRAS.456..485S} to radiate energy into its hosting cell with a spectral energy distribution (SED) according to particle mass, metallicity and age. 
For the luminosity of the Pop III.1 star we approximate its SED as a blackbody.
As discussed by \citet{2024arXiv241201828T}, a supermassive Pop III.1 star near the zero age main sequence structure is expected to have a photospheric temperature approaching $\sim10^5\:\rm{K}$ and have a total luminosity near that of the Eddington luminosity. In this scenario, the H-ionizing photon production rate is $\sim 10^{53}\,{\rm s}^{-1}$. We adopt this as a fiducial value for the Pop III.1 star, i.e.,  $Q_{\rm H}=10^{53}\,{\rm s}^{-1}$. 
The lifetime in this phase is expected to be at least $\sim3\,\rm{Myr}$, providing a conservative lower limit. If supported by residual heating from WIMP annihilation, we anticipate that the lifetime could extend to $10\,\rm{Myr}$. In this study, we adopt a source lifetime of $10\,\rm{Myr}$. While we do not explicitly vary the Pop III.1 lifetime, we analyze the evolution of the associated HII region at multiple time snapshots, and explore the degeneracy between stellar lifetime and the ionizing photon emission rate. 


The radial extent of the HII region, $r_{\mathrm{HII}}$, surrounding an ionizing source, such as the Pop III.1 star, evolves, in principle, through three primary phases. Initially, during the \textit{R-type} phase, the ionization front expands rapidly through the gas on timescales faster than the ability of the gas to respond dynamically to the change of temperature and pressure. If the ionizing source exists for long enough, which, in fact, is generally not the case for Pop III.1 stars (see below), then the ionization front eventually slows down as it approaches the Str\"omgren radius, $r_S$, where the enclosed recombination rate and the source H-ionizing photon injection rate are equal. The Str\"omgren radius, assuming fully ionized conditions in the HII region with a density that is equal to that of the mean density of the IGM, is given by
\begin{eqnarray}
r_S & = & \left( \frac{3\,Q_{\rm H}}{4\pi\,\alpha^\mathrm{B}\, n^2_{\mathrm{H}}} \right)^{1/3} = 61.3\, Q_{53}^{1/3} T_{\rm 3e4}^{0.27} \left(\frac{n_{\rm H}}{n_{{\rm H},z=30}}\right)^{-2/3}{\rm kpc}  \nonumber \\
 & = & 1.90\, Q_{53}^{1/3} T_{\rm 3e4}^{0.27} \left(\frac{n_{\rm H}}{n_{{\rm H},z=30}}\right)^{-2/3} \left(\frac{1+z_{\rm form}}{31}\right)\:{\rm cMpc}  \nonumber \\
 & = & 1.90\, Q_{53}^{1/3} T_{\rm 3e4}^{0.27} \left(\frac{1+z_{\rm form}}{31}\right)^{-1}\:{\rm cMpc},
\label{eq:rS}
\end{eqnarray}
where: $Q_{\rm H}$ is the Hydrogen-ionizing photon injection rate from the source and $Q_{53}=Q_{\rm H}/10^{53}\:{\rm s}^{-1}$; $n_{\mathrm{H}}$ is the gas number density of H nuclei; and $\alpha^\mathrm{B}=1.08 \times 10^{-13} T_{\rm 3e4}^{-0.8}\:{\rm cm}^3\:{\rm s}^{-1}$
is the case B hydrogen recombination coefficient, with normalization to a fiducial temperature of $3\times10^4\,\mathrm{K}$. 
In the final line of this equation, the evolution of the gas density has been absorbed into the redshift evolution term. 
%
Finally, in the \textit{D-type} phase, $r_{\mathrm{HII}}$ increases as the pressurized gas in the HII region expands into the surrounding medium.


The timescale to establish the HII region to the scale of $r_S$, i.e., the duration of the {\it R-type} phase, is $t_{\rm ion}\simeq (4/3)\pi r_S^3 n_{\rm H}/Q_{\rm H} = (\alpha^{B}n_{\rm H})^{-1} = 51.3[(1+z_{\rm form})/31]^{-3}\:$Myr, where we have adopted the mean density of the IGM for the formation redshift of $z_{\rm form}=30$. Thus the extent of the {\it R-type} HII region is likely to be set by the lifetime of the Pop III.1 star. For $t_*=10\:$Myr, the radius of the {\it R-type} front is given by \citep[see][]{2024arXiv241201828T}:

\begin{eqnarray}
r_R & = & \left(\frac{3\,t_*Q_{\rm H}}{4\pi n_{\rm H}}\right)^{1/3} = 35.5\, t_{*,10}^{1/3} Q_{53}^{1/3} \left(\frac{n_{\rm H}}{n_{{\rm H},z=30}}\right)^{-1/3}\:{\rm kpc}\label{eq:RRproper} \nonumber\\
 & = & 1.10\, t_{*,10}^{1/3} Q_{53}^{1/3} \left(\frac{n_{\rm H}}{n_{{\rm H},z=30}}\right)^{-1/3} \frac{(1+z_{\rm form})}{31}\:{\rm cMpc}\nonumber\\
 & = & 1.10\, t_{*,10}^{1/3} Q_{53}^{1/3} \:{\rm cMpc},
\end{eqnarray}
where $t_{*,10}=t_*/10\,\rm{Myr}$. As discussed by \citet{2024arXiv241201828T}, it is interesting that the analytic solution for the co-moving extent of the {\it R-type} front is independent of redshift. We see also from equation~\ref{eq:RRproper} that the typical expansion speed of the {\it R-type} front is $v_R = r_R /t_* \sim 0.12\, c$ for fiducial parameters and so is adequately captured by our choice of reduced speed of light approximation.

\textbf{Radiative cooling and heating processes:}
The hydrodynamical evolution of the gas is coupled to the local ionization via radiation pressure and the non-equilibrium hydrogen and helium thermochemistry, as described by \citet{rosdahl_ramses-rt_2013}.
%
%
In addition to primordial gas cooling, we account for metal-line cooling according to the gas metallicity, which gradually rises due to chemical enrichment following evolution of the standard stellar populations that form after the Pop III.1 source.
Above temperatures of $10^4\,\mathrm{K}$ we interpolate the pre-calculated tables of  {\sc cloudy} \citep{Ferland1998}, assuming photoionization equilibrium. 
%
Below $10^4\,\mathrm{K}$ we follow fine structure metal cooling rates from \citet{Rosen1995}.
 %
%





\textbf{Stellar feedback:}
While our focus in this paper is the radiative (ionizing) feedback from the Pop III.1 star, we also incorporate stellar feedback for the subsequent generations of stars.
This stellar feedback model includes radiation, as discussed earlier in this section, alongside momentum \citep{2014ApJ...788..121K} and 
magnetic energy \citep{martin-alvarez2021}.
Each stellar particle in our simulations corresponds to a single stellar population. 
The initial mass function (IMF) is modeled as a probability distribution function following \citet{Kroupa2001} and normalized over the complete range of masses.
%
%
%
%
The specific energy of each SN has a value 
$\varepsilon_\text{SN} = E_\text{SN} / M_\text{SN}$, where 
 $E_\text{SN} = 10^{51} \mathrm{erg}$ and $M_\text{SN} =10\, \mathrm{M}_{\odot}$.
Each supernova also returns a fraction of stellar mass back to the ISM.
We use $\eta_\text{SN} = 0.2$ for fraction of $M_\text{SN}$ returned as gas mass, and $\eta_\text{metal} = 0.075$ for the newly synthesized metals. 




\section{Results}\label{sec:res}

%
%
Our overall goal is to explore the evolution of the ionization front generated by Pop III.1 stars. In particular, the extent of the $\mathrm{HII}$ region is examined since it is proposed that this sets the isolation distance criteria of subsequent Pop III.1 progenitor halos, i.e., such halos would not form in regions that had been previously exposed to ionizing photons.
Ultimately, this allows us to estimate the number density of SMBHs that could originate from Pop III.1 progenitors.

%
\begin{figure}
    \centering
    \includegraphics[width=0.48\textwidth]{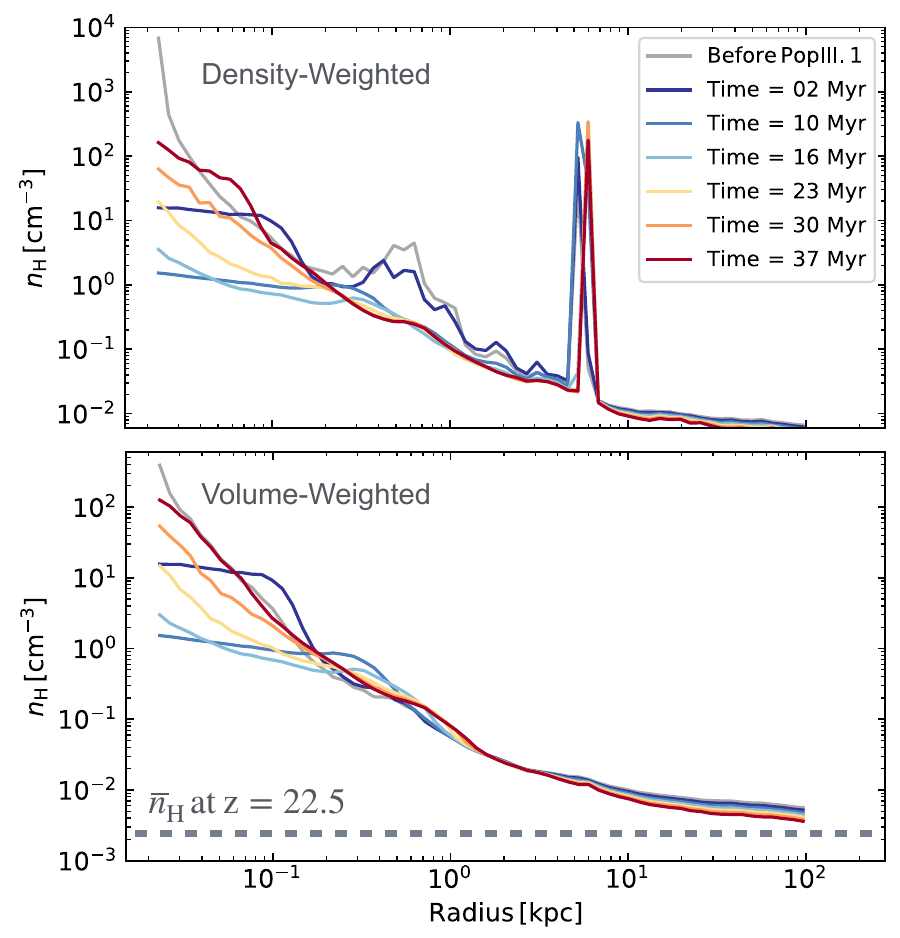}

    \caption{Density weighted (top) and volume weighted (bottom) gas radial profile before and after the formation of the Pop III.1 star.
    The plot shows the evolution of the gas number density within a radius of $100\,\rm{kpc}$ as a function of distance from the centre of the minihalo hosting the Pop III.1 star.
    Note that the secondary peak at $\sim$\,6 kpc corresponds to a neighboring minihalo.  
    The gray line represents the gas profile just before the formation of the Pop III.1 star, while the other colored lines show the profiles at various later times. 
    Before the Pop III.1 star forms, gas density in the centre of its host minihalo reaches $n_{\rm{H}} \simeq 10^{3}\,\rm{cm}^{-3}$, high enough to overcome the combined support of magnetic, thermal, and turbulent pressure, thereby satisfying the star formation criteria.
    After the Pop III.1 star forms, its radiation feedback heats the surrounding gas and pushes it outward, reducing the central density of the minihalo.  
    After the Pop III.1 star radiation ceases at $10\,\mathrm{Myr}$, gas in the central $\lesssim 60\,\mathrm{pc}$ falls back, increasing the central density.
    }
    \label{fig:density}
\end{figure}

\subsection{Evolution of primordial gas after Pop III.1 formation}\label{sec:popiii}

Figure.~\ref{fig:density} shows the evolution of the gas number density radial profile around the Pop III.1 star before and after its formation. The plot displays $n_{\rm{H}}(r)$ within a $\sim100\,\mathrm{kpc}$ radius from the source (proper distance). At $z_{\rm form}$ of $22.5$ this corresponds to about $2\,\rm{cMpc}$. Note that the average cosmic value of $n_{\rm H}$ at this redshift is $2.5\times 10^{-3}\:{\rm cm}^{-3}$ (dashed line).
The gray line represents the gas profile just prior to Pop III.1 star formation, while the other colored lines illustrate the profile at various times after this event.
The innermost region, within a $\mathrm{radius}\lesssim0.5\,\rm{kpc}$, corresponds to the gas profile inside the Pop III.1 host halo. A secondary peak in the gas density profile appears at a distance of $\sim6\,\mathrm{kpc}$, 
marking the location of a neighboring minihalo.
This neighboring halo has a virial mass of $7.5\times10^5\,M_\odot$, approximately $0.18$ times the Pop III.1 host halo virial mass.
%

Before the Pop III.1 star forms, its host halo has a virial mass of $\simeq4 \times 10^6\,M_{\odot}$ and contains a gas mass of $\simeq6\times10^5\,M_{\odot}$. In the centre of this minihalo, the gas number density reaches 
$n_{\rm{H}} \gtrsim 10^{3}\,\mathrm{cm}^{-3}$.
%
Upon the formation of the Pop III.1 star, a significant fraction of the gas in the minihalo, i.e., $\sim 1/6$, is removed and converted into the star. This removed gas also corresponds to the densest, innermost part of the minihalo.
Following the formation of the Pop III.1 star, its intense radiation heats the surrounding gas and drives it outward, further reducing the central density of the minihalo. 
When the radiation from the Pop III.1 star ceases after $10\,\mathrm{Myr}$, the gas in the centre of its host minihalo begins to cool and fall back inward, as indicated by the subsequent increase in gas density within the central $\lesssim 60\,\mathrm{pc}$.

\begin{figure*}
    \centering
    \includegraphics[width=0.85\textwidth]{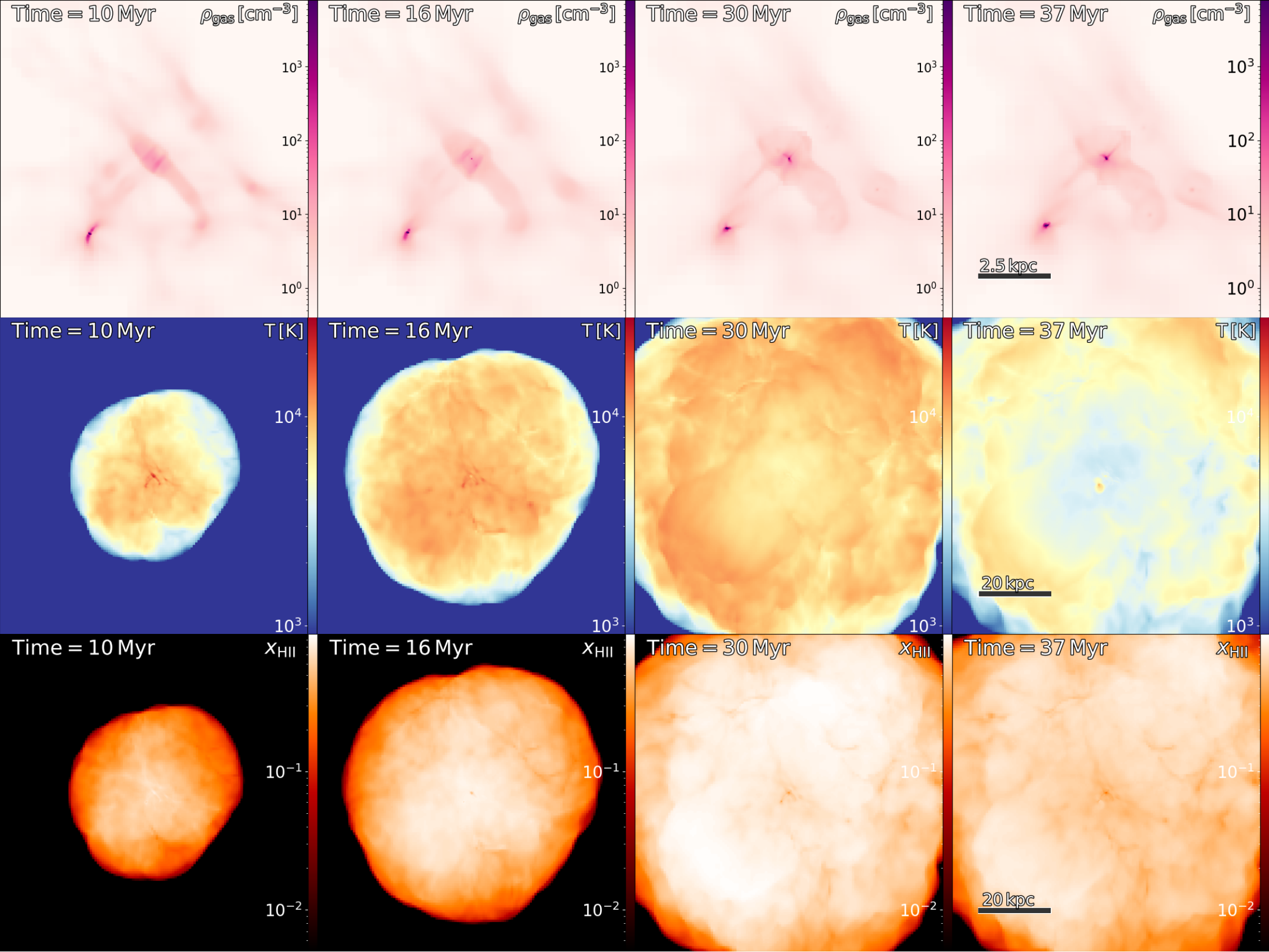}

    \caption{
    Density-weighted projections of gas number density (top row, on a scale of $10\,\rm{kpc}$ proper distance centered on the Pop III.1 source), temperature (middle row, on a scale of $80\,\rm{kpc}$ proper distance), and hydrogen ionized fraction  (bottom row, also on a scale of $80\,\rm{kpc}$ proper distance), from $10$ to $37\,\mathrm{Myr}$ (columns left to right) after the formation of the Pop III.1 star. Although radiation from Pop III.1 star ceases after $10\,\mathrm{Myr}$, the expansion of the ionized bubble continues, 
    altering the gas profile within a region approximately $80$ times the size of its host proto-galaxy.   
    Around $40\,\rm{Myr}$ after Pop III.1 formation, gas recombination begins in the dense regions at the centre of minihalos, where lower-mass stars subsequently start to form. 
    }

    \label{fig:T}
\end{figure*}

Figure~\ref{fig:T} shows snapshots from the time evolution of the ionized bubble created by the Pop III.1 star. 
%
It contains projected maps of gas number density in a central region of about $10\,\rm{kpc}$ (proper distance) on a side (top row), temperature in a wider region of about  $80\,\rm{kpc}$ (proper distance) on a side (middle row), and gas ionization fraction also of the wider region (bottom row), all weighted by gas density.
These maps illustrate how the ionized bubble created by the Pop III.1 star evolves over time.
The first column shows the state of the gas at $10\,\rm{Myr}$ after the formation of the star. As noted earlier, the star is assumed to collapse into a SMBH at this point. However, for this analysis, we use a simulation without the SMBH and any AGN phase in order to examine the effects of Pop III.1 radiation alone. The second, third and fourth columns show the gas evolution at later times, up to $37\,\rm{Myr}$, illustrating how the $\mathrm{HII}$ region continues to expand until about this time. With no continued injection of ionizing photons, the gas gradually undergoes recombination and becomes more neutral. Nevertheless, the relic $\mathrm{HII}$ region persists for at least a period of $\sim 30\,\rm{Myr}$, and the thermal imprint remains observable throughout. The time evolution of $r_{\rm{HII}}$ is presented in more detail in \S\ref{sec:L}.


\begin{figure*}
    \centering
    \includegraphics[width=\textwidth]{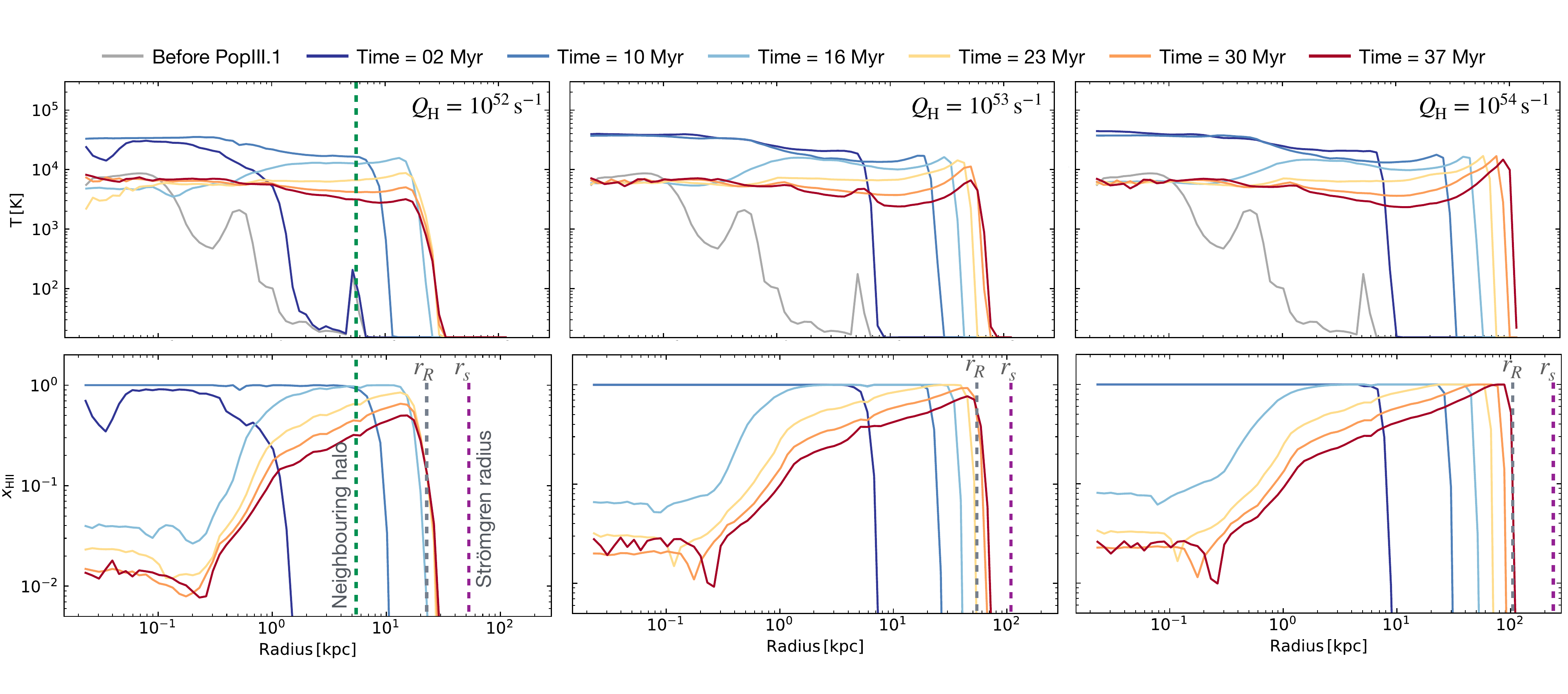}

    \caption{Profiles of temperature (top row) and hydrogen ionized fraction (bottom row) before and after Pop III.1 star formation. 
    Color codes are the same as in Fig.~\ref{fig:density}.
    The density profile is very similar in all models to the one in Fig.~\ref{fig:density}, which corresponds to $Q=10^{53}\,\rm{s}^{-1}$ in the middle panels.
    From left to right the H-ionizing photon injection rate of the Pop III.1 source increases from $Q=10^{52}$ to $10^{54}\,\rm{s}^{-1}$. The dashed purple line indicates the Strömgren radius for a Pop III.1 source with the corresponding radiation in each model. 
    Before the Pop III.1 star forms, the gas temperature is below $10^4\,\mathrm{K}$, cold enough to allow star formation.
    After its formation, radiation feedback from Pop III.1 raises the temperature to $\sim4$ times this initial value. It remains high until the ionized bubble expands, and the gas cools back. 
    The extent of the ionized bubble increases with the photon injection rate, and on average reaches $\sim80$ times the virial radius of the Pop III.1 host halo, defining the isolation distance for potential Pop III.1 star progenitor halos. 
    }
    \label{fig:gasProfilePopIII}
\end{figure*}

\subsection{Impact of Pop III.1 ionizing luminosity on HII region structure and evolution}\label{sec:L}

In the Pop III.1 SMBH seeding scenario, primordial metal-free gas that has been ionized is not expected to form Pop III.1 stars, because of the enhanced cooling and fragmentation that occurs under such conditions. Rather, collapsing minihalos impacted by ionization feedback are expected to form lower-mass Pop III.2 stars (see \S\ref{sec:intro}). For this reason it is important to characterize the extent of relic HII regions around Pop III.1 sources, since they may self-regulate the production of the entire SMBH population and set its global cosmic abundance. 

Figure~\ref{fig:gasProfilePopIII} shows the evolution of the gas temperature (top row) and hydrogen ionization fraction (bottom row) before and after the formation of Pop III.1 stars,  within a sphere of radius of about $100\,\mathrm{kpc}$ (proper distance). 
The $x$-axis represents the distance from the centre of the minihalo hosting the Pop III.1 star.
The colors of the lines are the same as in Fig.~\ref{fig:density}, where gray represents the gas profile just before the Pop III.1 star forms, and the other colors indicate different times, ranging from $2$ to $37\,\mathrm{Myr}$ after its formation. Each column corresponds to a different model, with the H-ionizing photon injection rate of the Pop III.1 star increasing from $Q_{\rm{H}} = 10^{52}$ to $10^{54}\,\rm{s}^{-1}$ from left to right. 
The position of the closest minihalo to the Pop III.1 host halo, at a distance of $\sim 6\,\mathrm{kpc}$, is marked by the green dashed line in the panels in the left column. 




Ionization fractions rise to near unity in an inner zone that effectively defines the size of the $\mathrm{HII}$ region while the star is shining. Then, after $10\,\rm{Myr}$, when the ionizing flux from the star drops to zero, ionization levels in the inner regions begin to fall. Since the recombination rate scales inversely with the density, this drop is fastest in the denser, innermost regions. 
However, during this period from $10$ to $\sim30\,\rm{Myr}$ the HII region still continues to expand. 
In the higher ionizing luminosity cases, the ionization fraction remains close to unity in the outer regions throughout the evolution to $37\,\rm{Myr}$, when expansion is still continuing. In the lower ionizing luminosity cases, the expansion stalls by $\sim 30\,\rm{Myr}$, after which the ionization fraction gradually reduces.
Although we do not directly explore variations in the Pop III.1 lifetime here, we infer from the different lines in Fig.~\ref{fig:gasProfilePopIII}, which show the $\mathrm{HII}$ region at different times, that changes in stellar lifetime are degenerate with variations in the ionizing photon emission rate.



The dashed purple line represents the analytically calculated Strömgren radius for a Pop III.1 source with the corresponding emission in each model (see Eq.~\ref{eq:rS}).  This calculation assumes spherical symmetry and a homogeneous gas distribution with a density evaluated to be that of the mean value of the IGM. This estimate also neglects the contributions of helium ions to the ionization and recombination balance.
%
This estimated Strömgren radius is approximately a factor of two larger than the final extent of the HII region, confirming that its evolution is ``{\it R-type} limited'' (see \S\ref{sec:method}). The predicted size from the analytical {\it R-type} expansion given by eq.~\ref{eq:RRproper} is shown by the gray dashed lines. Despite its simplifications, the {\it R-type} expansion provides a more accurate description of $r_{\rm{HII}}$ than the Str\"omgren radius. These analytic estimates are compared with the actual maximum relic HII region sizes in Table~\ref{tab:params}. 
%
%
As the ionizing luminosity increases in the sequence from $Q_{\rm{H}} =10^{52}$ to $10^{53}$ to $10^{54}\:{\rm s}^{-1}$, $r_{\rm{HII}}$ increases from $\simeq29$ to $74$ to $116\,\rm{kpc}$ (proper distance). From $Q_{\rm{H}} =10^{52}$ to $10^{53}\:{\rm s}^{-1}$ this growth roughly follows the expected scaling of $r_R\propto Q_{\rm H}^{1/3}$. At the highest ionizing luminosity, the scaling becomes slightly shallower, which can be attributed to the fact that at $37\,\mathrm{Myr}$, the ionization front in the $Q_{\rm H}=10^{54}\:{\rm s}^{-1}$ case is still expanding. Therefore, its final extent is expected to exceed $\simeq116\,\mathrm{kpc}$.

The temperature of the gas responds to its ionization state. Before the Pop III.1 star forms, the gas temperature is below $10^4\,\mathrm{K}$.
Following Pop III.1 star formation, its radiative feedback heats the surrounding gas, raising the temperature to about $3.5\times10^4\,\rm{K}$ during the first $10\,\rm{Myr}$ of evolution. 
After death of the Pop III.1 star, temperatures gradually decrease, but remain at several thousand K at the final timestep, when the ionized bubble reaches its maximum extent.
Once the bubble expands sufficiently, the gas cools back to its original temperature.

The extent of the ionized region indicates the zone in which further Pop III.1 star formation is not expected to occur. In the $Q_{\rm H}=10^{53}\:{\rm s}^{-1}$ case, this is about $50\,\rm{kpc}$ proper distance for a region that is nearly fully ionized. The region that experiences about $1\%$ ionization fraction is about $74\,\rm{pc}$ (i.e., the size $r_{\rm HII}$ reported in Table~\ref{tab:params}). 
However, one expects that pre-existing minihalos, being overdense structures, would maintain lower ionization fractions than seen in the diffuse IGM and so could potentially form new Pop III.1 stars if located within these HII region boundaries. 
Thus, for more precise estimates of the feedback zone around a Pop III.1 source that prevents further Pop III.1 star formation, a more detailed tracking of marginally irradiated minihalos is needed.


\begin{table*}
\caption{\small
  Parameters varying in the simulation runs and the resulting extent of the HII region. 
  Columns are as follows: 
  1) Model ID. 
  2) Pop III.1 ionizing radiation photon rate. 
  3) Environment of Pop III.1 host halo.
  4) Redshift of Pop III.1 formation.
  5) Average gas number density in the IGM at the time of Pop III.1 formation.
  6) Stromgren radius generated by Pop III.1 ionizing radiation.
  7) {\it R-type} expansion generated by Pop III.1 ionizing radiation.
  8) Radius of ionized bubble generated by Pop III.1 star in proper units.
  9) Radius of ionized bubble generated by Pop III.1 star in comoving units.
  \label{tab:params}}
  \centering
\begin{tabular}{l c c c c c c c c}

 \hline
      {\texttt{ID}} &  {$Q_{\rm{H}}$} &  {Environment} &  {$z_{\rm{form}}$} & $n_{\mathrm{H}}$ & {$r_{\mathrm{S}}$}  & {$r_{\mathrm{R}}$}  &{$r_{\mathrm{ HII}}$} & {$r_{\mathrm{ HII}}$}\\
       &  $[\rm{s}^{-1}]$ &   &  & [$10^{-3}\,\mathrm{cm^{-3}}$]& $[\rm{kpc}]$ & $[\rm{kpc}]$ & $[\rm{kpc}]$ & $[\rm{cMpc}]$\\
 \hline
 {\texttt{Avrg52}} &  {$10^{52}$} &  {Average-density} &  {22.5} & $2.50$ & $52.51$ & $23.10$ &{29.3} & 0.61\\
 {\texttt{Avrg53}} &  {$10^{53}$} &  {Average-density} &  {22.5} &  $2.50$ & $106.46$ & 46.83&{74.0} & 1.49\\  
  {\texttt{Avrg54}} &  {$10^{54}$} &  {Average-density} &  {22.5} &  $2.50$ &  $229.36$ & $100.89$&{115.6} & 2.30\\ 
  \hline
 {\texttt{Film53}} &  {$10^{53}$} &  {Over-dense} &  {29.8}  & $5.63$ & $61.90$ & $35.71$& {49.8} & 1.29\\
 {\texttt{Void53}} &  {$10^{53}$} &  {Under-dense} &  {19.8} & $1.73$ & $135.95$ & $52.92$&{74.0} &  1.35\\




  \hline

\end{tabular}
\end{table*}

\begin{figure}
    \centering
    \includegraphics[width=0.5\textwidth]{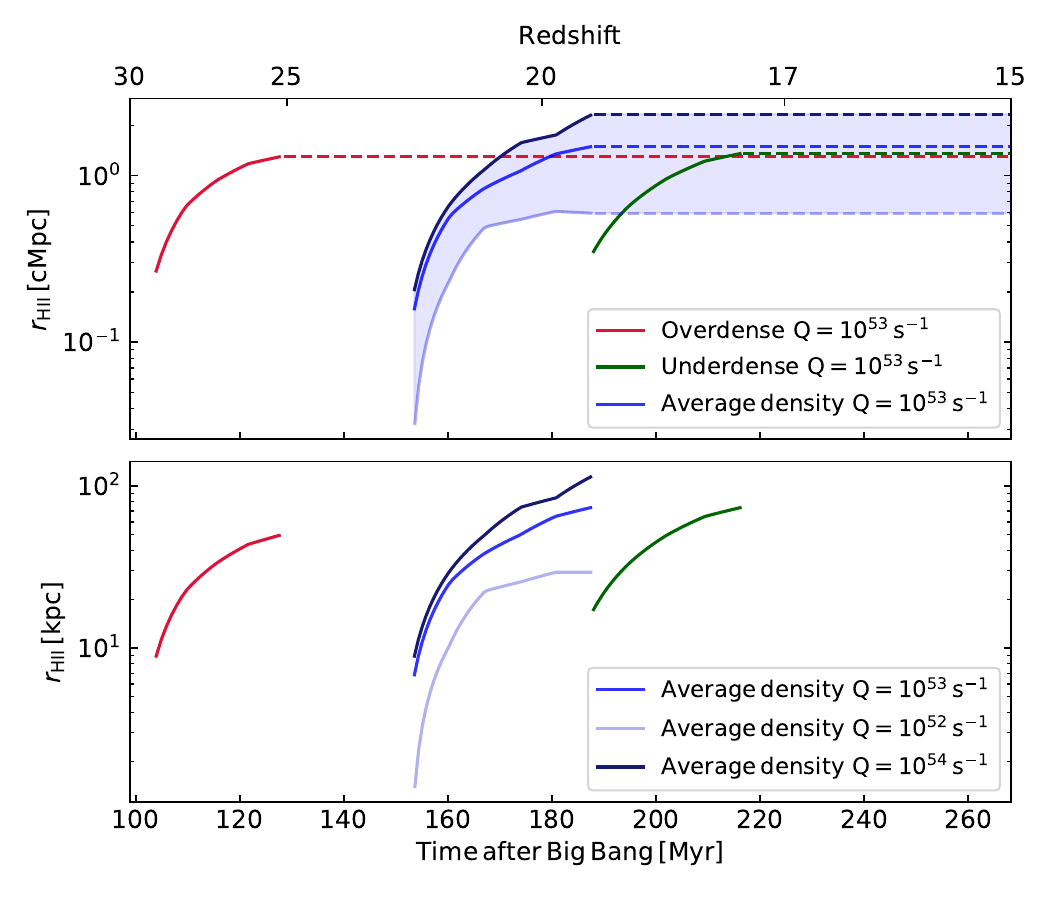}
    \caption{
    Evolution of the HII region radius ($r_{\mathrm{HII}}$) in terms of comoving (top) and proper (bottom) distance as a function of time for the three different environments: an overdense
    , an average-density (blue), and an underdense (green) model 
    The solid lines show the evolution of $r_{\mathrm{HII}}$ for $\sim40\,\rm{Myr}$ after Pop III.1 star formation, which marks the period before significant gas recombination begins.
    The dashed lines show the extrapolation of $r_{\mathrm{HII}}$ up to redshift $z=15$. 
    For the normal region, the shaded area shows the range spanned by lower ($10^{52}\,\mathrm{s}^{-1}$) and higher ($10^{54}\,\mathrm{s}^{-1}$) photon injection rates. 
    }
    \label{fig:rHII}
\end{figure}

\subsection{Impact of environment on HII region structure and evolution}\label{sec:env}

Here, we focus on how the environment in which the Pop III.1 host minihalo forms affects the structure and evolution of the HII region.
%
%
To do this, we adopt a common approach of varying the local value of $\sigma_8$ in the initial conditions of our simulations while keeping the mean density unchanged \citep[e.g.,][]{2020ApJS..250....2V}. This directly impacts the clustering strength and timing of structure formation.
We model three different environments by adjusting the local value of $\sigma_8$ from its fiducial value of $\sigma_{8,\,\mathrm{local}}= 0.8159$:
increasing it by $0.29$ to enhance clustering and produce a filament-intersection environment, and decreasing it by $0.08$, to reduce clustering and produce a void-like environment.
These values are derived from a parent $100\,\mathrm{Mpc}$ DMO simulation by randomly sampling approximately $1000$ spheres of radius $8\,\rm{h^{-1}\,cMpc}$ and computing the matter overdensity $\delta_m$ in each sphere. 
While the standard deviation of the resulting overdensity distribution matches the fiducial value of $\sigma_8$, the filament-intersection and void-like regions are selected from spheres at the one sigma tails of this distribution.
We then rerun the zoom-in simulations using the same $7.55\,\mathrm{cMpc}$ simulation box as in the fiducial case discussed in the previous section, but with varying the $\sigma_{8,\,\mathrm{local}}$ value in the initial conditions (see \S\ref{sec:method} for details). 
This approach allows us to simulate the Pop III.1 hosting halo in high-, average- and low-density regions in a controlled and consistent way. 
The redshift at which the Pop III.1 star forms in each model, and the average hydrogen number density in the IGM at the corresponding redshift, are listed in Table~\ref{tab:params}. 
%


In Figure~\ref{fig:rHII} we illustrate the extent of $r_{\mathrm{HII}}$ in both comoving (top) and proper (bottom) units as a function of time, for Pop III.1 stars formed in these different cosmic environments. 
Beside a photon injection rate of $Q_{\rm{H}} = 10^{53}\,\mathrm{s}^{-1}$ in all models, the shaded area represents the range spanned by lower ($Q_{\rm{H}} = 10^{52}\,\mathrm{s}^{-1}$) and higher ($Q_{\rm{H}} = 10^{54}\,\mathrm{s}^{-1}$) photon injection rates in the average-density region.
The solid lines show the evolution of $r_{\mathrm{HII}}$ for $\sim40\,\rm{Myr}$ after Pop III.1 star formation, which marks the period before significant gas recombination begins. The method used to compute $r_{\mathrm{HII}}$ is the same as that described in \S\ref{sec:L}.
The dashed lines show the extrapolation of $r_{\mathrm{HII}}$ down to redshift $z=15$, enabling a comparison between the impact of changing $Q_{\rm{H}}$ values and environmental conditions.


The timing of Pop III.1 star formation varies across environments. 
Pop III.1 formation follows the same star formation prescription as less massive stellar particles in the simulation (see \S\ref{sec:method} for details). 
The formation redshift corresponds to the moment when gravitational collapse overcomes magneto-thermal-turbulent pressure, resulting in the formation of the first star in the minihalo.
In the fiducial model (average-density region), the Pop III.1 star forms at $z = 22.5$. 
In the 
overdense model, where gravitational collapse occurs earlier, the star forms at a higher redshift ($z = 29.8$). 
Conversely, in the 
underdense model, where gas collapse is slower, Pop III.1 formation is delayed until $z = 19.8$.
These differences in formation redshift, combined with the higher gas density at earlier times, result in variations in the extent of the ionization front. 
For a Pop III.1 star forming in the overdense model $r_{\mathrm{HII}}$ reaches $\sim50\,\rm{kpc}$ proper distance, while it extends to $ \sim74\,\rm{kpc}$ proper distance in an underdense or average-density region. Note that the comoving extent of $r_{\mathrm{HII}}$ converges to $\simeq1\,\rm{cMpc}$ across different models for the Pop III.1 environmental conditions and is approximately independent of the redshift of Pop III.1 formation. 
%
%

Table~\ref{tab:params} summarizes all the models examined, highlighting the impact of both the environment and photon injection rate. 
While the environment influences the growth and final size of the ionized front, variations in the photon injection rate have a more significant impact, as shown by the wider range of $r_{\mathrm{HII}}$ for different $Q_{\rm{H}}$ rates in Fig.~\ref{fig:rHII}.
\begin{figure*}
    \centering
    \includegraphics[width=0.85\textwidth]{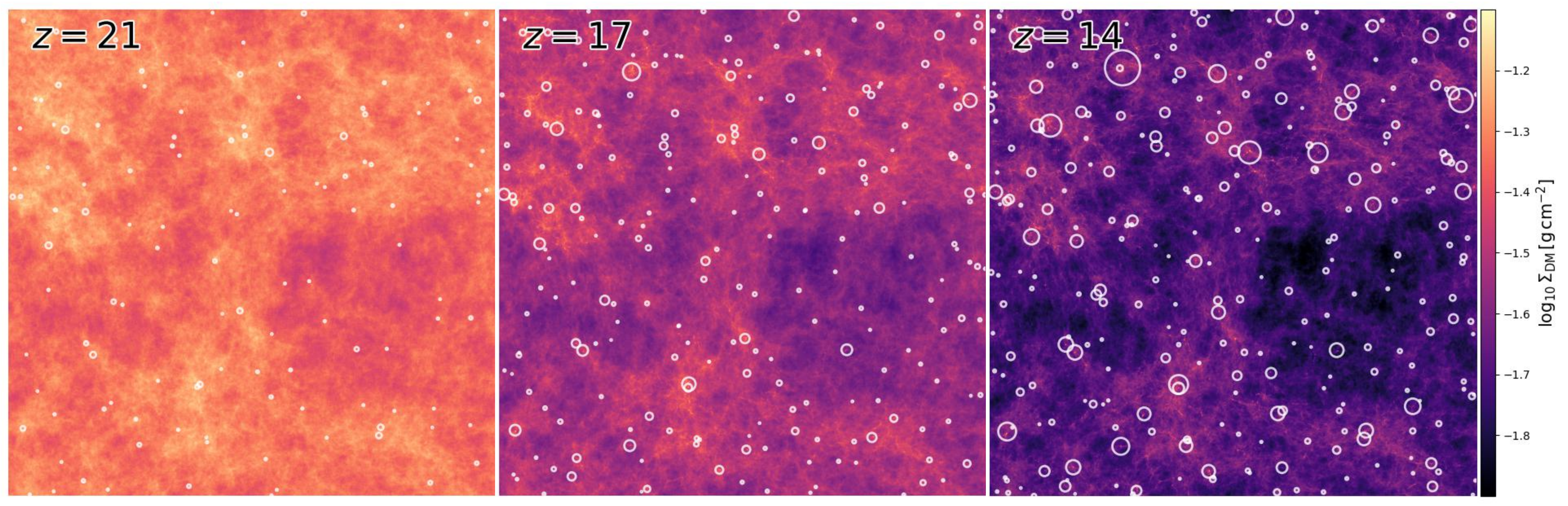}

    \caption{Dark matter density projection in a $11\,\mathrm{Mpc}$ box at redshifts $z\sim21$, $17$, and $14$. Seeded halos are marked with white circles, with the circle size scaled to the halo mass. 
    To select halos eligible for SMBH seeding, we first select those with masses above $10^6\,M_{\odot}$, and then apply an isolation criterion based on the comoving distance $r_{\mathrm{ HII}}=1\,\mathrm{cMpc}$ to remove neighboring halos (see text for details). This ensures that the most massive halos are seeded before their neighbors. The volume 
    shows the spatial distribution of seeded halos across redshifts. As redshift decreases, the number of massive halos hosting Pop III.1 seeded SMBHs increases, 
    and new low-mass halos fulfilling isolation criteria shift to the edges of voids, away from larger halos.
}
    \label{fig:DMO}
\end{figure*}

\subsection{SMBHs number density}\label{sec:nSMB}

\begin{table*}
\caption{\small
  Parameters varying in the DMO simulation runs. 
  Columns are as follows: 
  1) Box size, $L_{\mathrm{Box}}$. 
  2) redshift, $z$. 
  3) Cube root of the total number of dark matter particles in the simulation box, $\mathrm{Np}^{1/3}$.
  4) Mass resolution of dark matter particles, $\mathrm{Mp}$.
  5) Minimum mass of halos detectable with $100$ particles by the halo finder,  $\mathrm{Mh_{min}}$.
  6) Maximum mass of halos detectable with $100$ particles by the halo finder, $\mathrm{Mh_{max}}$.
  7) Total number of halos more massive $10^6\,M_{\odot}$, $\mathrm{Nh}$.
  8) Total number of halos seeded by SMBHs, $\mathrm{Nh_{seeded}}$.
  \label{tab:dmo_boxes}}
  \centering
\begin{tabular}{c c c c c c c c}

 \hline
    $L_{\mathrm{Box}}\,\mathrm{[cMpc]}$&  z  & $\mathrm{Np}^{1/3}$ & $\mathrm{Mp}\,[10^4\,M_{\odot}]$  & $\mathrm{Mh_{min}}\,[M_{\odot}]$  & $\mathrm{Mh_{max}}\,[M_{\odot}]$    &  $\mathrm{Nh}$ & $\mathrm{Nh_{seeded}}$ \\
 \hline
    $11$ & $21$  & $1024$  & $4.97$ &   $5.02\times10^6$ & $1.16\times10^8$   & $1104$ & $214$\\
    
    $11$ & $17$  & $1024$  & $4.97$ &   $5.02\times10^6$ & $5.65\times10^7$   & $11623$ & $513$\\
    
    $11$ & $14$  & $1024$  & $4.97$ &   $5.02\times10^6$ & $2.88\times10^9$   & $39839$ & $698$\\

    $7$ & $21$  & $1024$  & $1.22$ &   $1.23\times10^6$ & $5.65\times10^7$   & $2922$ & $140$\\
    
    $7$ & $17$  & $1024$  & $1.22$ &   $1.23\times10^6$ & $3.02\times10^8$   & $21220$ & $201$\\
    
    $7$ & $14$  & $1024$  & $1.22$ &   $1.23\times10^6$ & $1.10\times10^9$   & $57637$ & $233$\\

  \hline

\end{tabular}
\end{table*}

In this section we investigate the number density of halos expected to host SMBHs originating from Pop III.1 star progenitors.
For this, we use two parent DMO simulation boxes with volumes of $(11\,\mathrm{Mpc})^3$ and $(7\,\mathrm{Mpc})^3$, each resolved with $1024^3$ particles. The mass resolutions are approximately $5\times10^4\,M_{\odot}$ and $1.2\times10^4\,M_{\odot}$ per particle, respectively.
These resolutions are chosen to well-resolve the halo mass function down to minihalos of $\sim 10^6\,M_{\odot}$ (with at least $100$ particles per halo), while also capturing the distribution of more massive halos, up to $\simeq10^9\,M_{\odot}$, within the simulation volumes.  
In other words, the larger box samples rarer, higher-mass halos while the smaller, higher resolution box better resolves the minihalo population and the low-mass end of the halo mass function. The convergence between the two boxes is described in more detail in Appendix~\ref{sec:app}. 

%
%

There are two key parameters to determine which halos are eligible for SMBHs seeding.
The first parameter is the 
mass of the minihalos. High-resolution simulations have shown that a minimum halo mass of about $10^6\,M_{\odot}$ is needed for primordial composition gas to be able to cool and contract to high density as a requirement for Pop III star formation \citep[e.g.,][]{bromm2002,abel_formation_2002,2024arXiv241107282K}.
Thus, using the \textsc{adaptahop} halo finder, we select all halos with masses above this threshold.
For the halo identification, we employ a friend-of-friend (FOF) algorithm with a linking length defined as $\Delta x(b) = b\times\overline{l}$, where $b=0.2$ is the dimensionless linking length parameter and $\overline{l}=L_{\mathrm{Box}}/N_{\mathrm{p}}^{1/3}$ represents the mean inter-particle separation in the simulation. The linking length thus sets the maximum allowed distance between particles for them to be considered part of the same halo.
The second parameter is the isolation distance of halos from their neighbors. Guided by the models of \S\ref{sec:env}, we set this isolation distance to be $d_{\rm iso}=r_{\rm HII}=1.0\,\rm{cMpc}$, keeping this value constant with redshift. 
%
%
In brief, to identify halos eligible for SMBHs seeding, for each simulation snapshot, we first select all halos with masses greater than $10^6\,M_{\odot}$. 
We then apply a filtering procedure starting from the most massive halo and progressively moving down the mass hierarchy. For each halo, beginning with the most massive, we exclude all neighboring halos within a distance of 
$r_{\mathrm{HII}} =1\,\mathrm{cMpc}$,
from the list of potential seed candidates. 
This approach ensures that massive and sufficiently isolated halos remain eligible to host a Pop III.1 star before their lower-mass neighbors.

Figure~\ref{fig:DMO} shows the projection of dark matter surface density in the $(11\,\mathrm{Mpc})^3$ simulation box, at three different redshifts, ranging from $z=21$ to $14$. 
As time progresses, the number of halos with masses above $10^6\,M_{\odot}$ increases and more massive halos begin to emerge as smaller minihalos grow.
The halos hosting SMBHs are marked with white circles, with the radii of circles scaling with halo mass.
The apparent overlap of circles arises from projection effects, where halos separated along the line of sight (z-axis) appear superimposed in the x–y plane.
The volume displayed contains regions of different densities. 
At high redshifts, SMBHs hosting halos are mostly confined to dense regions. These eventually evolve to become more massive halos. 
As redshift decreases, the number of seeded halos at the high-mass end increases, while the population of seeded halos at the low-mass end remains relatively constant.
Most newly seeded low-mass halos at low redshifts tend to appear near the edges of voids, in environments isolated from the radiation field expected around more massive neighbors.

\begin{figure}
    \centering
    \includegraphics[width=0.45\textwidth]{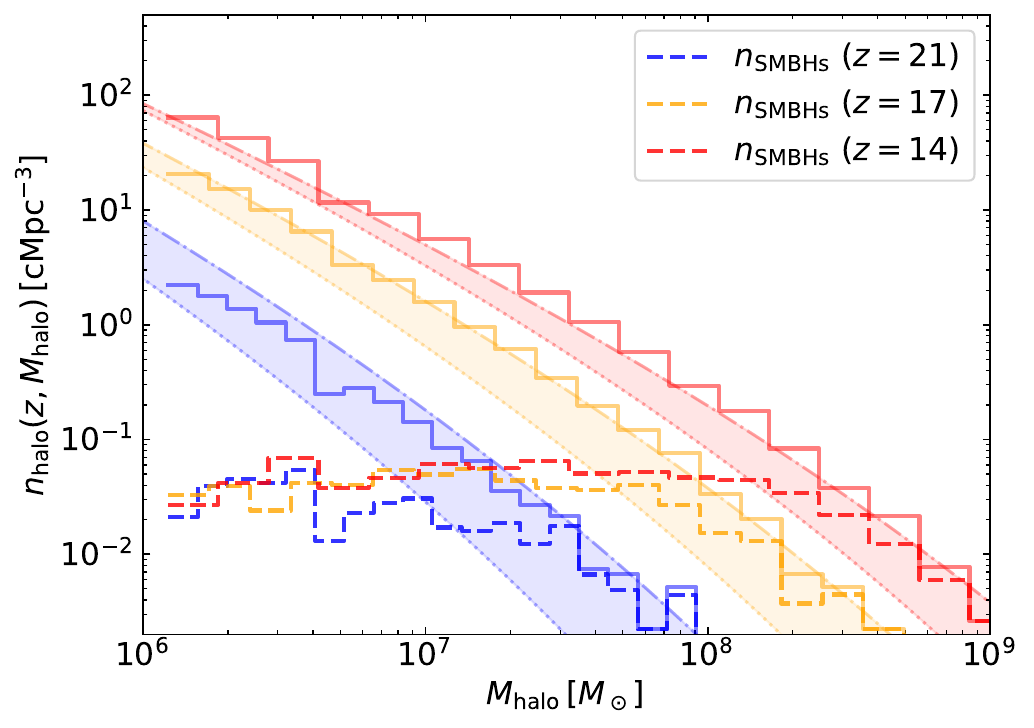}

    \caption{Halo 
    mass function at different redshifts. Solid histogram lines represent the number density of halos with $M_{\rm{halo}}\gtrsim10^6\,M_{\odot}$. 
    The number density in each mass bin, $n_{\mathrm{halo}}(z,\,M_{\rm{halo}})$,
    is calculated by combining 
    the halo counts
    from both $7$ and $11\,\rm{Mpc}$ simulation boxes and dividing by their respective volumes. Dashed histogram lines show the number density of seeded halos, separated by a 
    distance of $\simeq1\,\rm{cMpc}$.
    The colored curves show for comparison, the halo number density fits predicted by the  \citet{1974ApJ...187..425P} and \citet{2001MNRAS.323....1S} halo mass functions at redshifts $z=21$ (blue), $z=17$ (yellow), and $z=14$ (red).
    }
    \label{fig:nSMBH_hist}
\end{figure}

The resulting number density as a function of halo mass for SMBHs emerging from Pop III.1 progenitors is shown in Figure~\ref{fig:nSMBH_hist} with dashed lines. Different colored lines correspond to different redshifts, $z=21$ (blue), $z=17$ (yellow), and $z=14$ (red).
The solid histogram lines show the number density of halos with halo mass $M_{\rm{halo}}\geq10^6\,M_{\odot}$, extracted from the DMO simulations with box sizes $7$ and $11\,\rm{Mpc}$. 
To combine the halo populations from the two simulation volumes, we adopt the patching method described in \citet{obrennan_halo_2024}. 
To provide a theoretical comparison reference, we show as dot-dashed and dotted lines the predictions for the halo mass function by \citet{1974ApJ...187..425P} (PS) and \citet{2001MNRAS.323....1S} (SMT), respectively.
The PS formalism models halo formation using a Gaussian probability distribution for the threshold density required for spherical collapse. The SMT mass function follows a similar approach but improves halo abundance predictions by accounting for ellipsoidal collapse.
The shaded area between the PS and SMT predictions represents the range of expected number densities at redshifts: $z=21,\, 17$ and $z=14$.


The dashed histogram lines show the subset of halos, that satisfy the SMBHs seeding criteria. 
As redshift decreases, increasingly massive halos begin to appear.
However, not all of them are seeded with SMBHs. This is because the most massive halos typically form in dense environments, clustered with many nearby halos. 
For example, at redshift $z=21$, a seeded halo in the $11\,\rm{Mpc}$ and $7\,\rm{Mpc}$ boxes has up to $69$ and $220$ neighbors, respectively. 
Due to the isolation criterion, none of these neighboring halos are eligible for seeding, as they fall within the exclusion radius of $1\,\rm{cMpc}$. As a result, the number of seeded halos grows more slowly than the total number of halos with $M_{\rm{halo}} \gtrsim 10^6\,M_{\odot}$.
Most of these neighbors are low-mass minihalos, which is why the histogram flattens at the low-mass end.


%

\begin{figure}
    \centering
    \includegraphics[width=0.45\textwidth]{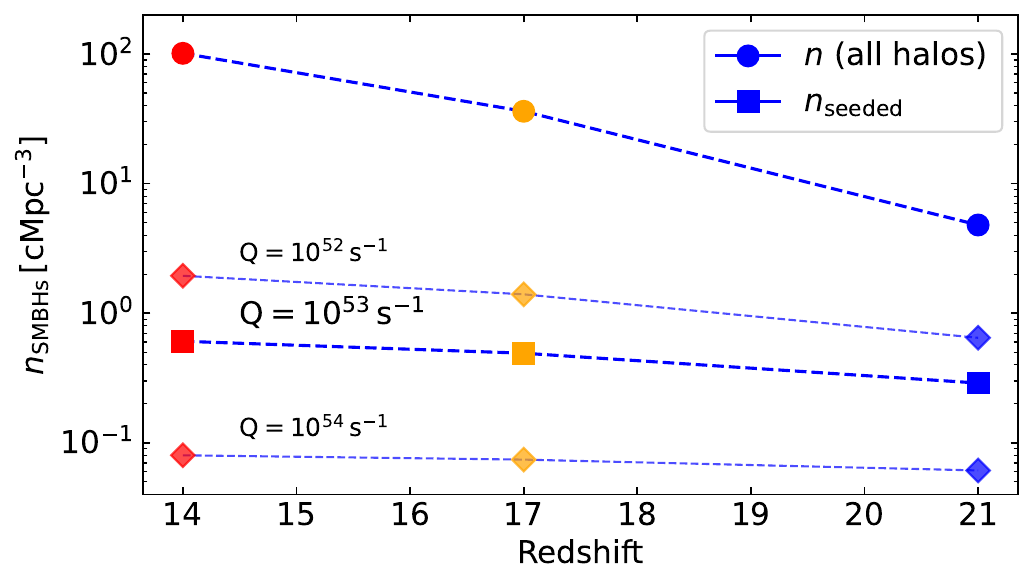}

    \caption{Number density of PopIII.1 progenitor SMBHs, $n_{\mathrm{SMBHs}}$, as a function of redshift. Circles show the number density of all halos with $M_{\rm{halo}}\gtrsim10^6\,M_{\odot}$,
    calculated as the average of the number densities of halos summed across both the $7$ and $11\,\rm{Mpc}$ simulation boxes for each mass bin.
    Squares show the number density of all seeded halos, separated by a distance of $\simeq1\,\rm{cMpc}$ in the fiducial model with $Q = 10^{53}\,\rm{s^{-1}}$. 
    Diamonds show the number density of seeded halos obtained for separation distances of $0.6$ and $2.3\,\rm{cMpc}$, corresponding to the wide range of H-ionizing photon emission rates explored ($Q = 10^{52}$ - $10^{54}\,\rm{s^{-1}}$). Poisson errors are smaller than the symbol sizes and thus not visible. 
        }
    \label{fig:nSMBH_z}
\end{figure}

Figure~\ref{fig:nSMBH_z} shows the evolution of the number density of SMBHs, $n_{\mathrm{SMBHs}}$, as a function of redshift. The circles represent the number density of all halos with $M_{\rm{halo}} \gtrsim 10^6\,M_{\odot}$, obtained by summing $n_{\mathrm{halo}}(z,\,M_{\rm{halo}})$ across all mass bins at each redshift from the histogram in Fig.~\ref{fig:nSMBH_hist}. For this, the total halo counts from the $7$ and $11\,\rm{Mpc}$ DMO simulation boxes are combined \footnote{Number density of halos, $n_{\mathrm{halo}}$, is computed as the average of $\Sigma N_{\mathrm{halo},i}(z,\,M_{\rm{halo}})/V_i$ across both simulation volumes $V_i$, where $N_{\mathrm{halo},i}(z,\,M_{\rm{halo}})$ is the number of halos in each mass bin $M_{\rm{halo}}$ at redshift $z$. The same method is used to compute the number density of seeded halos, $n_{\mathrm{SMBHs}}$.}. 
%
Squares indicate the number density of halos that have been seeded with SMBHs. 
At redshift $z=21$, the number density of halos in the $11\,\rm{Mpc}$ box is $\simeq9\,\rm{cMpc}^{-3}$, with $\simeq\,0.42\,\rm{cMpc}^{-3}$ of those halos seeded, resulting in a seeded fraction of $\sim5\%$. In the $7\,\rm{Mpc}$ box, the corresponding number density is $\simeq0.8\,\rm{cMpc}^{-3}$ with $\simeq0.16\,\rm{cMpc}^{-3}$ seeded halos, corresponding to a seeded fraction of $\sim20\%$. The figure shows the average of both boxes combined, which gives a number density of $\simeq5\,\rm{cMpc}^{-3}$ for halos above the mass threshold, with $\simeq0.3\,\rm{cMpc}^{-3}$ of them seeded, resulting in a seeded fraction of $\sim6\%$.
It is worth mentioning that the resulting $n_\mathrm{SMBHs}$ obtained here agrees in order of magnitude with the analytical prediction for the number density of SMBHs driven by HII region feedback during the R-type expansion phase of Pop III.1 stars, as presented in \citet{2024arXiv241201828T}, where $n_{\mathrm{SMBHs}} = {3}/{4\pi r_{\mathrm{HII}}^3}\simeq0.18\,\mathrm{cMpc}^{-3}$. This value is derived using the same parameters for calculating $r_{\mathrm{HII}}$ as those adopted in the \texttt{Avrg53} model presented in this study.

In Fig.~\ref{fig:nSMBH_z}, as the number of halos with masses above $10^6\,M_\odot$ increases over time, the number of seeded halos, and consequently $n_{\mathrm{SMBH}}$ also rises, from an initial $\sim0.3\,\rm{cMpc}^{-3}$ at redshift $z= 21$ to $0.6\,\rm{cMpc}^{-3}$ at the ending redshift $z= 14$.
However, the increase in the number density of seeded halos is less steep than that of all halos. As previously discussed, this is because more halos form at lower redshift, but they typically appear in clusters with other halos, and therefore excluded by the isolation criterion.
We note that this conservative method provides a lower limit on the number of halos eligible for seeding.
In contrast, the total number of massive halos without applying any isolation criterion sets an upper limit. 
The actual number of seeded halos is expected to fall between these two bounds.
Nevertheless, even with this conservative seeding method, our result of $n_{\mathrm{SMBH}} \gtrsim 10^{-1},\mathrm{cMpc}^{-3}$ by $z = 14$ is consistent with the high number densities implied by recent James Webb Space Telescope observations \citep{2024ApJ...964...39G, scholtz_net-zero_2024, 2024ApJ...963..129M, 2025arXiv250305537I}.
%

\section{Conclusions and discussion}\label{sec:sum}

%
%
In this work, we have investigated the seeding 
of SMBHs in the Pop III.1 scenario for the first time using cosmological radiation–hydrodynamical simulations.
In this scenario, SMBHs originate from the collapse of massive, isolated Pop III.1 stars that form in pristine, undisturbed minihalos.
During their lifetimes, these stars evolve to have a phase of intense H-ionizing photon emission, with rates of $\sim 10^{53}\,\rm{s}^{-1}$ sustained for $\sim 10^7\,\rm{yr}$. 
This powerful radiative feedback generates large HII regions that extend into the IGM, establishing an isolation distance criterion for Pop III.1 hosting halos. Upon their collapse, these stars leave behind black holes of comparable mass, i.e. $\simeq 10^5\,M_\odot$, which accrete from their surroundings and appear as a nascent AGN, offering a compelling explanation for the abundance of SMBHs observed at high redshift. Our main findings are as follows:
%


    %





\begin{itemize}


    \item The ionizing emission from the Pop III.1 source impacts a large cosmological volume, extending significantly beyond the halo. The ionized bubble expands to reach from $30$ up to $100$ times the virial radius as we vary the ionizing photon rate from $Q_{\rm{H}} = 10^{52}$ to $10^{54}\,\rm{s}^{-1}$. This ionized sphere size variation sets the characteristic exclusion length, which serves as the minimum separation required for halos to be capable of forming SMBHs from Pop III.1 progenitors.

    \item We explore three different environments for the Pop III.1 host halo: an over-dense region resembling cosmic filaments, an average-density region, and an under-dense region similar to cosmic voids. 
    The main effect of environmental clustering variations is to shift the formation redshift of Pop III.1 sources, with over-dense regions accelerating their formation ($z\sim30$), and under-dense ones delaying it further ($z\sim20$). 
    \item When comparing the importance of different effects on the extent of the isolation distance, we find variations on the ionizing radiation budget to dominate over environmental effects and their influence on formation times.
    \item 
    We find that HII region feedback in the \textit{R-type} expansion phase reaches a nearly redshift-independent comoving radius of $\sim1\,\mathrm{cMpc}$, corresponding to an average physical distance of approximately $70\,\rm{kpc}$ across models with varying formation redshifts.
    This result aligns with the predictions from the Pop III.1 semi-analytic framework of \citet{2025MNRAS.536..851C} finding a preference for $\lesssim75\,\rm{kpc}$ physical isolation distance.
    
%
    %


    \item We estimate the number density of SMBHs at high redshifts using two seeding criteria: a halo mass threshold of $10^6\,M_{\odot}$ and a minimum separation of 
    $\sim 1\,\mathrm{cMpc}$,  corresponding to the extent of the HII region ionized by a PopIII.1 source. Applying these criteria, we predict a number density of SMBHs, $n_{\mathrm{SMBH}}\gtrsim10^{-1}\,\mathrm{cMpc}^{-3}$ by $z=14$, consistent with the high number densities recently observed by the {\it James Webb Space Telescope}.

\end{itemize}


Overall, our work characterizes the approximate uncertainty range of the ionization impact by exploring how variations in Pop III.1 source properties and their environments affect the extent of the ionized region.
Using high-resolution simulations and comprehensive physics modeling, we obtain self-consistent predictions for the number density of SMBHs based on the seeding of halos $\gtrsim10^6\,M_{\odot}$ at redshifts $z\lesssim30$ and the exclusion criteria described above. 
In this model, we assume efficient formation of supermassive stars, such that all minihalos meeting the seeding criteria, namely, exceeding the mass threshold and being isolated from neighboring ionizing sources, form a Pop III.1 star, and serve as progenitor halos for SMBHs.
%
Our conclusions remain valid for alternative seeding mechanisms that yield similar BH seed masses.
In future work, we will use a larger set of simulations to investigate more diverse galaxy populations in the Pop III.1 SMBH seeding scenario.

\section*{Acknowledgements}
M.S. acknowledges the support from the Swiss National Science Foundation under Grant No. P500PT\_214488 and the Chalmers Initiative on Cosmic Origins (CICO) postdoctoral fellowship.
J.C.T. acknowledges support from ERC Advanced Grant 788829 (MSTAR) and the
CCA Sabbatical Visiting Researcher program.
This work used the DiRAC@Durham facility managed by the Institute for Computational Cosmology on behalf of the STFC DiRAC HPC Facility (www.dirac.ac.uk). The equipment was funded by BEIS capital funding via STFC capital grants ST/P002293/1, ST/R002371/1 and ST/S002502/1, Durham University and STFC operations grant ST/R000832/1. DiRAC is part of the National e-Infrastructure. This work was performed using resources provided by the Cambridge Service for Data Driven Discovery (CSD3) operated by the University of Cambridge Research Computing Service (www.csd3.cam.ac.uk), provided by Dell EMC and Intel using Tier-2 funding from the Engineering and Physical Sciences Research Council (capital grant EP/P020259/1), and DiRAC funding from the Science and Technology Facilities Council (www.dirac.ac.uk).
P.M. acknowledges support by the Italian Research centre on High Performance Computing Big Data and Quantum Computing (ICSC) and by a PRIN 2022 PNRR project (code no. P202259YAF) funded by ``European Union – Next Generation EU''. B.W.K. acknowledges support from program HST-AR-17547 provided by NASA through a grant from the Space Telescope Science Institute, which is operated by the Associations of Universities for Research in Astronomy, Incorporated, under NASA contract NAS5-26555.
S.M.A. is supported by a Kavli Institute for Particle Astrophysics and Cosmology (KIPAC) Fellowship, and by the NASA/DLR Stratospheric Observatory for Infrared Astronomy (SOFIA) under the 08\_0012 Program. SOFIA is jointly operated by the Universities Space Research Association, Inc. (USRA), under NASA contract NNA17BF53C, and the Deutsches SOFIA Institut (DSI) under DLR contract 50OK0901 to the University of Stuttgart. V.C. thanks the ERC BlackHoleWeather project and PI Prof. Gaspari for salary support.

\section*{DATA AVAILABILITY}
The data employed in this manuscript is to be shared upon reasonable request contacting the corresponding author.

\bibliographystyle{mnras}
\bibliography{bibliography, references}

\appendix
\section{Halo mass function}\label{sec:app}

\begin{figure}
    \centering
    \includegraphics[width=0.48\textwidth]{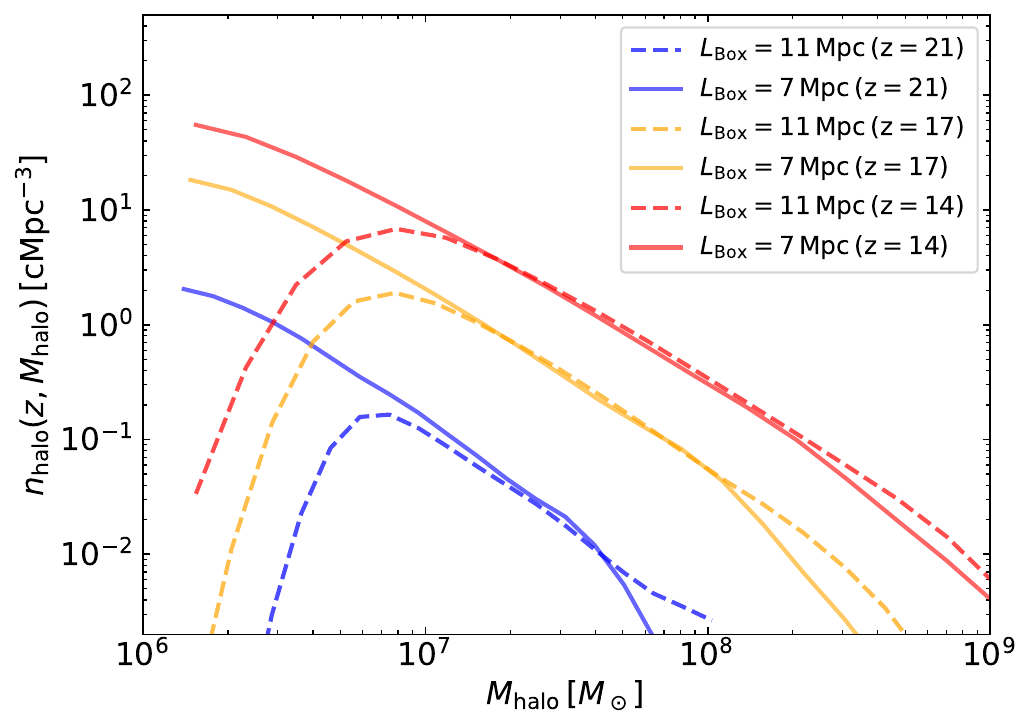}

    \caption{Halo mass function at different redshifts for the $7$ and $11\,\rm{Mpc}$ simulation boxes, shown with dashed and solid lines, respectively. The number density in each mass bin, $n_{\mathrm{halo}}(z,\,M_{\rm{halo}})$,
    is calculated as the number of halos divided by the volume of the respective simulation box. Different colors represent redshifts: $z=21$ (blue), $z=17$ (yellow), and $z=14$ (red). The smaller box better resolves minihalos at the low-mass end, while the larger box captures the more massive halos.}  
    
    \label{fig:HMF}
\end{figure}

Figure.~\ref{fig:HMF} shows the halo mass function in two DMO simulation boxes with comoving side lengths of $L_{\mathrm{Box}} =11\,\mathrm{Mpc}$ (dashed lines) and $L_{\mathrm{Box}} = 7 \,\mathrm{Mpc}$ (solid lines). Different colors represents different redshifts, ranging from $z=21$ to $z=14$. The particle mass resolution of $5\times10^4\,M_{\odot}$ and $1.2\times10^4\,M_{\odot}$ in two boxes allows us to resolve minihalos with at least $100$ particles per halo down to $\sim 10^6\,M_{\odot}$, while also sampling more massive halos up to $\sim 10^9\,M_{\odot}$. 
The smaller, higher resolution box better resolves the minihalo population and the low-mass end of the halo mass function, while the larger box captures more massive halos. 
The number density of halos with masses of a few $10^7\,M_{\odot}$ convergences between the two boxes at $z=21$. As structure growth progresses, the mass range over which the halo mass function converges widens from $M_{\mathrm{halo}}\simeq10^7$ to $10^8\,M_{\odot}$. By redshift $z=14$, halos spanning a broad mass range are present in both simulation boxes.
The combined number densities in Fig.~\ref{fig:nSMBH_hist} are computed by averaging over both simulation volumes as $\Sigma N_{\mathrm{halo},i}(z,\,M_{\rm{halo}})/V_i$ using non-zero halo counts $N_{\mathrm{halo},i}(z,\,M_{\rm{halo}})$ in each mass bin $M_{\rm{halo}}$.
This combined approach enables us to sample the full halo mass range relevant for Pop III.1 black hole seeding.

\bsp	
\label{lastpage}
\end{document}